\documentclass[aps,prl,preprint,superscriptaddress]{revtex4-1}
\usepackage{graphicx}
\usepackage{amssymb}
\usepackage{color}
\usepackage[version=3]{mhchem}

\begin{document}

%\preprint{}

\title{Magnetic and magnetoresistive behavior of the ferromagnetic heavy fermion \ce{YbNi_2}}

\author{O. Olic\'on}
%\thanks{}
\affiliation{Instituto de Investigaciones en Materiales, Universidad Nacional Aut\'onoma de M\'exico, Ciudad de M\'exico, 04510, M\'exico}
\author{R. Escamilla}
\affiliation{Instituto de Investigaciones en Materiales, Universidad Nacional Aut\'onoma de M\'exico, Ciudad de M\'exico, 04510, M\'exico}
\author{A. Conde-Gallardo}
\affiliation{Departamento de F\'isica, Centro de Investigaci\'on y de Estudios Avanzados del Instituto Polit\'ecnico Nacional, A.P. 14740, C.P. 07300, Ciudad de M\'exico, M\'exico}
\author{F. Morales}
\affiliation{Instituto de Investigaciones en Materiales, Universidad Nacional Aut\'onoma de M\'exico, Ciudad de M\'exico, 04510, M\'exico}

\date{\today}

\begin{abstract}
We present a study on the magnetic susceptibility $\chi(T)$ and
electrical resistance, as a function of temperature and magnetic
field $R(T,H)$, of the ferromagnetic heavy fermion \ce{YbNi_2}. The
X-ray diffraction analysis shows that the synthesized
polycrystalline samples crystallizes in the cubic Laves phase
structure C15, with a spatial group $Fd\overline{3}m$. The magnetic
measurements indicate a ferromagnetic behavior with transition
temperature at 9 K. The electrical resistance is metallic-like at
high temperatures and no signature of Kondo effect was observed. In
the ferromagnetic state, the electrical resistance can be justified
by electron-magnon scattering considering the existence of an energy
gap in the magnonic spectrum. The energy gap was determined for
various applied magnetic fields. Magnetoresistance as a function of
applied magnetic field, subtracted from the $R(T,H)$ curves at
several temperatures, is negative from 2 K until about 40 K for all
applied magnetic fields. The negative magnetoresistance originates
from the suppression of magnetic disorder by the magnetic field.
\end{abstract}

%\pacs{}
\keywords{Heavy fermion \and \ce{YbNi_2} alloy \and Ferromagnetism
\and Magnetoresistance}

\maketitle

\section{Introduction}
\label{intro}

The Yb and Ce-based alloys and compounds have been a subject of
interest due to its physical properties like heavy fermion behavior,
mixed valence state and Kondo lattice behavior
\cite{Stewartd-f,HF,StewartHF,BauerCeYb,MValence}. These properties
are associated to the hybridization between conduction-electron band
and $f$-electron band. The interaction is mediated via the
polarization of the conduction electrons, which is known as RKKY
(Ruderman-Kittel-Kasuya-Yosida) interaction \cite{RKKY1,RKKY2}. In
the last years the study of quantum criticality in heavy fermion
(HF) systems have constituted a subject of interest in the condensed
matter physics, due to the  phenomena related to magnetic critical
points at low temperatures, where quantum fluctuations compete with
classical thermal fluctuations \cite{QFT1,QFT2}. Examples of such
phenomena are unconventional superconductivity and non-Fermi liquid
(NFL) behavior. Usually, the search for the existence of a quantum
critical point (QCP) is through the modification of the system using
a non-thermal parameter, such as pressure,  magnetic field or
chemical doping.

The ferromagnetic (FM) order in HF systems is rare compared with the
antiferromagnetic (AFM) order or superconductivity. In the Laves
phases the $\rm{RNi_2}$ alloys, where R is a rare-earth element, the
magnetic properties are associated to the rare-earth element. It was
proposed that in these alloys the $d$-shell of Ni atoms is full,
$3d^{10}$ configuration \cite{skrabek,RNi2}, consequently Ni does
not carry a magnetic moment and does not contribute to the magnetic
properties. The exchange interactions responsible of magnetism must
take place between the $4f$ electron magnetic moments of the
rare-earth element mediated by the conduction electrons.

In particular, the $\rm{YbNi_2}$ alloy has been classified as a
heavy fermion because the Sommerfeld coefficient determined at low
temperatures is $\gamma=573$ mJ/mol K \cite{Rojas}. At low
temperatures \ce{YbNi_2} is ferromagnetic below a transition
temperature $T_C$=10.5 K \cite{Rojas}. Calculations of the
electronic density of states show that the electrons of the
conduction band are hybridized with the Yb $4f$-electrons, this
hybridization confirms the heavy fermion character of \ce{YbNi_2}
\cite{yan14}. However, there is a lack of studies on the electronic
transport properties. The present work studies the magnetic field
effects on the electrical resistance of the \ce{YbNi_2} alloy. We
found that in the ferromagnetic state the temperature dependence of
the electrical resistance is produced by the scattering of the
conduction electrons by magnons. Furthermore, an energy gap in the
magnonic spectrum was inferred.

\section{Experimental details}

Polycrystalline samples of \ce{YbNi_2}, were synthesized weighting
stoichiometry quantities of high purity Yb (99.9\% Sigma-Aldrich)
and Ni (99.99\% Sigma-Aldrich) powders. An excess of Yb 10\% wt was
added to compensate the Yb loss because its low melting point. The
powders were mixed and pelletized into a plastic bag with Ar
atmosphere. Afterward, the pellet was melted in high-purity Ar
atmosphere in an arc furnace. The produced button was turned and
remelted several times in order to obtain an homogeneous sample. The
structural characterization was performed with the X-ray powder
method using a  Bruker diffractometer model D8 Advanced, with a
Cu-$\rm{K_{\alpha}}$ radiation. The diffraction patterns were
collected at room temperature and over a $2\theta$ range of
$20-120^{\circ}$ with a step size $0.02^{\circ}$. The Rietveld
refinement of the X-ray pattern was performed using the MAUD
software \cite{maud}. DC-magnetization measurements were made in a
SQUID based magnetometer (Quantum Design, MPMS-5) in the temperature
range of 2-300 K with an applied magnetic field of 200 Oe.
Electrical resistance, as a function of temperature and magnetic
field, was measured through the conventional AC four-probe method.
For this purpose the sample was connected using Cu wires glued with
Ag paint. The measurements were performed in a Physical Property
Measurement System (PPMS, Quantum Design) at temperatures ranging
from 2 K to 300 K and magnetic fields ranging from 0 up to 90 kOe.
The magnetic field was applied perpendicular to the electrical
current applied to the sample.

\begin{figure}[t]
\begin{center}
\includegraphics[width=0.48\textwidth]{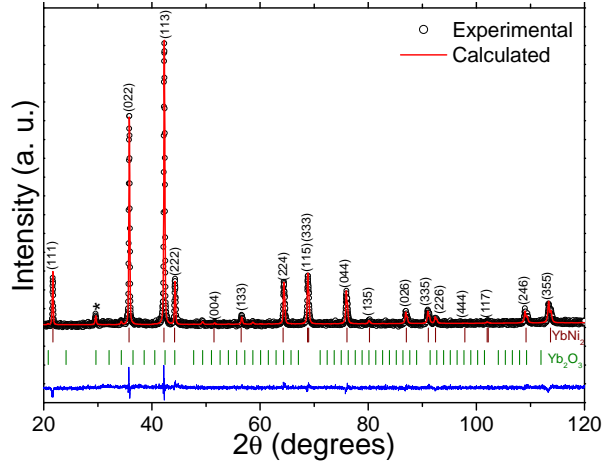}
\caption{\label{difra} X-ray diffraction pattern and Rietveld fitting of \ce{YbNi_2}. The $R$-factors of refinement are: $R_{wp}=5.2,\,R_b=3.9,\,R_{exp}=4.1$ and $\sigma^2=1.2$. The vertical red bars indicate the Bragg reflections of \ce{YbNi_2}. The continuous line at the bottom is the difference between the experimental data and the fit.}
\end{center}
\end{figure}

\section{Results and discussion}

Figure \ref{difra} shows the powder X-ray diffraction pattern of
\ce{YbNi_2}. The analysis of these data indicates that the
crystalline structure of the sample corresponds to \ce{YbNi_2}
(JCPDS No 03-065-5017), however a faint trace of \ce{Yb_2O_3} (JCPDS
No 00-041-1106) are observed. This oxide is indicated by an asterisk
in the diffraction pattern. The X-ray diffraction pattern was
Rietveld-fitted using a space group $Fd\overline{3}m$ (No 227),
considering the presence of \ce{Yb_2O_3}. The value obtained for the
lattice parameter $a=7.0965(3)$ {\AA} is in agreement with the
values reported previously \cite{buschow72,Palenzona,Rojas}.

\begin{figure}[ht]
\begin{center}
\includegraphics[width=0.48\textwidth]{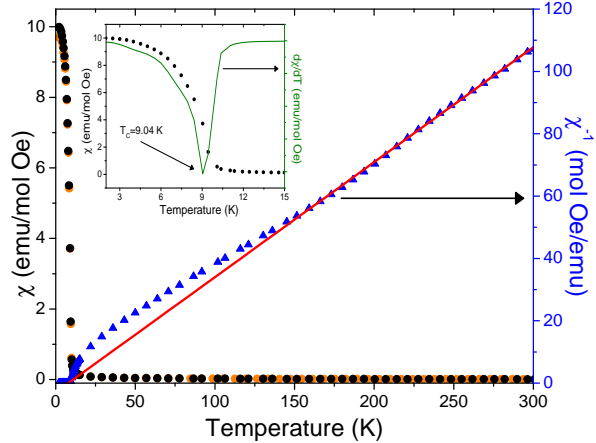}
\caption{\label{Mag} Magnetic susceptibility $\chi=M/H$ (circles) and $\chi^{-1}$ (triangles), as a function of temperature measured under 200 Oe. The continuous line is a linear fit of $\chi^{-1}(T)$ data from 150 K to room temperature. The inset shows the determination of $T_C$ like the minimum in $d\chi/dT$.}
\end{center}
\end{figure}

Figure \ref{Mag} shows the temperature dependence of the magnetic
susceptibility, $\chi(T)$, and the inverse magnetic susceptibility,
$\chi^{-1}(T)$, measured in a magnetic field of 200 Oe. The magnetic
susceptibility between 2 K and 300 K shows a fast increase at low
temperatures, which characterize the ferromagnetic behavior. The
inset in this figure shows the $\chi(T)$ and $d\chi/dT$ curves at
low temperatures, here the ferromagnetic behavior is clearly noted
and the minimum in $d\chi/dT$ is defined as the ferromagnetic
transition temperature $T_C=9$ K. This value is lower than $T_C=
10.5$ K earlier reported \cite{ivanshin14,Rojas}. In spite of the
tiny amount of \ce{Yb_2O_3} in our sample, its influence on the
magnetic properties of \ce{YbNi_2} was not considered, since it has
an antiferromagnetic transition temperature at 2.1 K \cite{adachi}.

To analyze the paramagnetic state, $\chi^{-1}(T)$ is plotted in Fig.
\ref{Mag} (right scale). The continuous line is a linear fit of the
data from 150 K to 300 K. The linear behavior can be described by
the Curie-Weiss law, $\chi(T)=C/(T-\theta_{CW})$, where $C$ is the
Curie constant and $\theta_{CW}$ is the Curie-Weiss temperature. The
effective magnetic moment, $\mu_{eff}$, can be determined from $C$.
In the present case, the values of these parameters obtained from
the fitting were;  $\theta_{CW} = 7.86$ K and $C=2.71$ emu K/mol Oe
which gives $\mu_{eff}=4.65 \mu_B$. This effective magnetic moment
is near to the theoretical value, $\mu_{eff}=g_J[J(J+1)]^{1/2}=4.53$
$\mu_B$, reported for the Yb$^{3+}$ ion \cite{Blundell}.

Rojas {\it et al.} \cite{Rojas} calculated the magnetic entropy of
\ce{YbNi_2} based on specific heat measurements. The entropy
saturates around 130 K, which is congruent with a full population of
the crystalline electric field (CEF) effects. As observed in Fig.
\ref{Mag}, there is a deviation of $\chi^{-1}(T)$ from the linear
behavior, this deviation could be related to the CEF effect on the
magnetization below 150 K.

\begin{figure}[t]
\begin{center}
\includegraphics[width=0.48\textwidth]{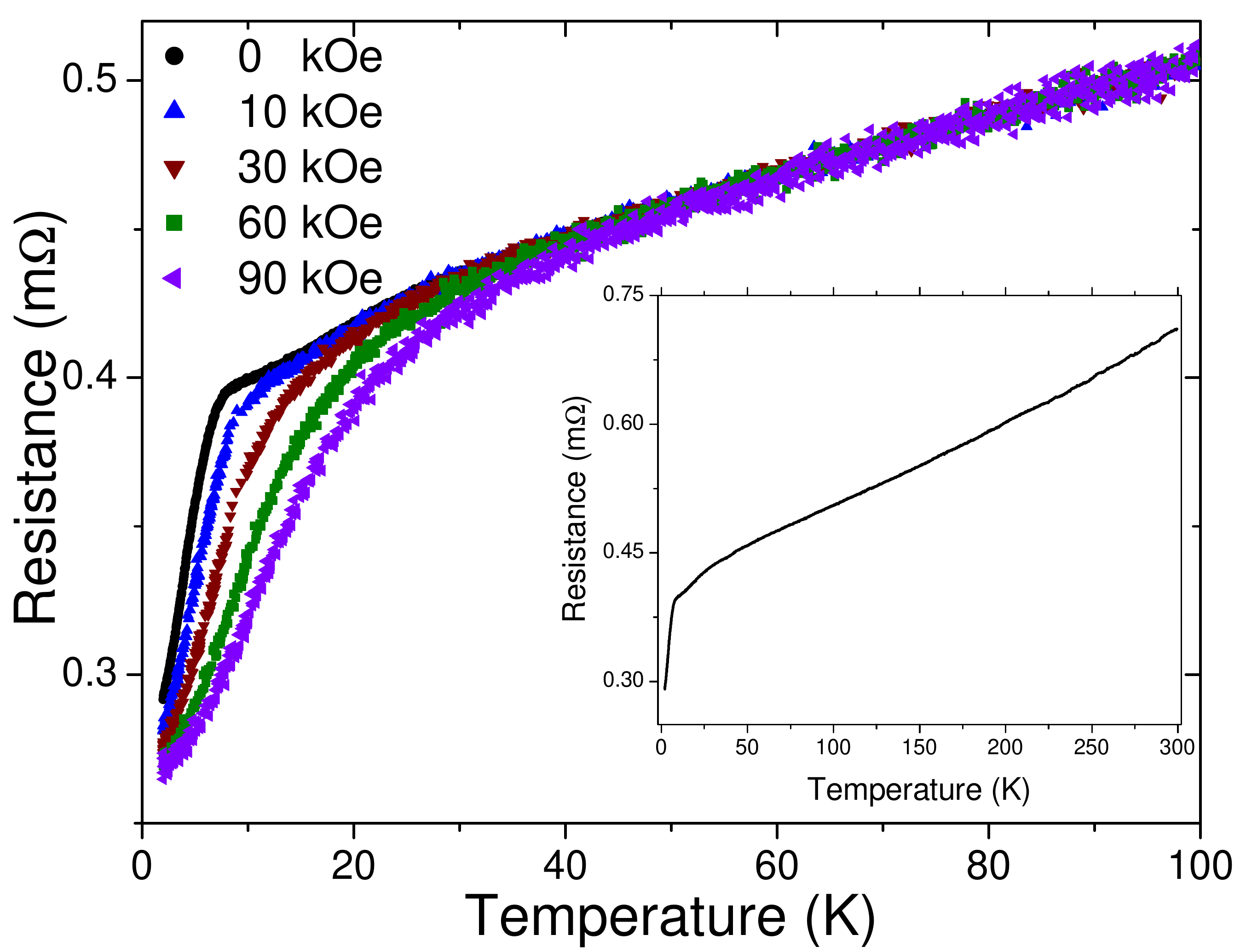}
\caption{\label{RTB} Electrical resistance as a function of temperature and applied magnetic field, measured between 2 K and 100 K. The inset shows the electrical resistance, without applied magnetic field, at temperatures between 2 K and 300 K.}
\end{center}
\end{figure}

The electrical resistance of \ce{YbNi_2} as a function of
temperature in zero magnetic field is shown in the inset of Fig.
\ref{RTB}. The curve displays a metallic-like behavior and shows a
sharp decrease about the ferromagnetic transition at low
temperatures. However, for temperatures between $T_C$ and 15 K the
electrical resistance shows a $T^2$ dependence indicative of a Fermi
liquid behavior. The onset of the sharp resistance decrease is about
$T_{on}=8$ K, this value is near to the $\theta_{CW}$ determined
from the magnetic measurements. It is noteworthy the absence of a
logarithmic increase of $R(T)$, signature of a Kondo effect, this
behavior suggest that the Kondo effect is weak and that the RKKY
interaction dominates. The Kondo temperature has been reported as
$T_K=27$ K, determined from the jump of the specific heat at the
ferromagnetic transition \cite{Rojas}. Taking the approximation
$|\theta_{CW}|=2T_K$ \cite{blanco94,yamauchi00} we obtain $T_K=3.93$
K. This $T_K$ value is lower than reported, but congruent to the
Kondo signature fault in $R(T)$.

The residual resistance ratio ($RRR$) is defined by the relation
$RRR=R$(300 K)$/R$(2 K), this ratio provides qualitative information
about the electron scattering by structural defects. For \ce{YbNi_2} was
obtained $RRR=$2.7, which indicates that our samples
have defects, probably included Yb vacancies because of the low
melting point of Yb. The $RRR$ value is similar to the already
observed values in other $\rm{RNi_2}$ alloys
\cite{RRR-RNi2,gratz96}.

The main panel of Fig. \ref{RTB} shows the $R(T)$ curves measured
under different applied magnetic fields. Above 40 K the curves are
similar to the one measured at zero magnetic field. The changes
occur at low temperature, the electrical resistance decreases as the
magnetic field increases. At first sight, the application of a
magnetic field smooth out the magnetic transition and moves up in
temperature as the magnetic field is increased; this behavior is
characteristic of a ferromagnetic ordering. The reduction of $R(T)$
with magnetic field suggest a negative magnetoresistance $(MR)$, as
expected for a ferromagnetically ordered system. Qualitatively, it is observed
that the $MR$ magnitude increases as the temperature
increases from 2 K, it reaches a maximum and goes to zero as the
temperature increases. A negative MR indicates that the magnetic
fluctuations decreases as the magnetic field increases
\cite{Moriya}.

\begin{figure}[ht]
\begin{center}
\includegraphics[width=0.48\textwidth]{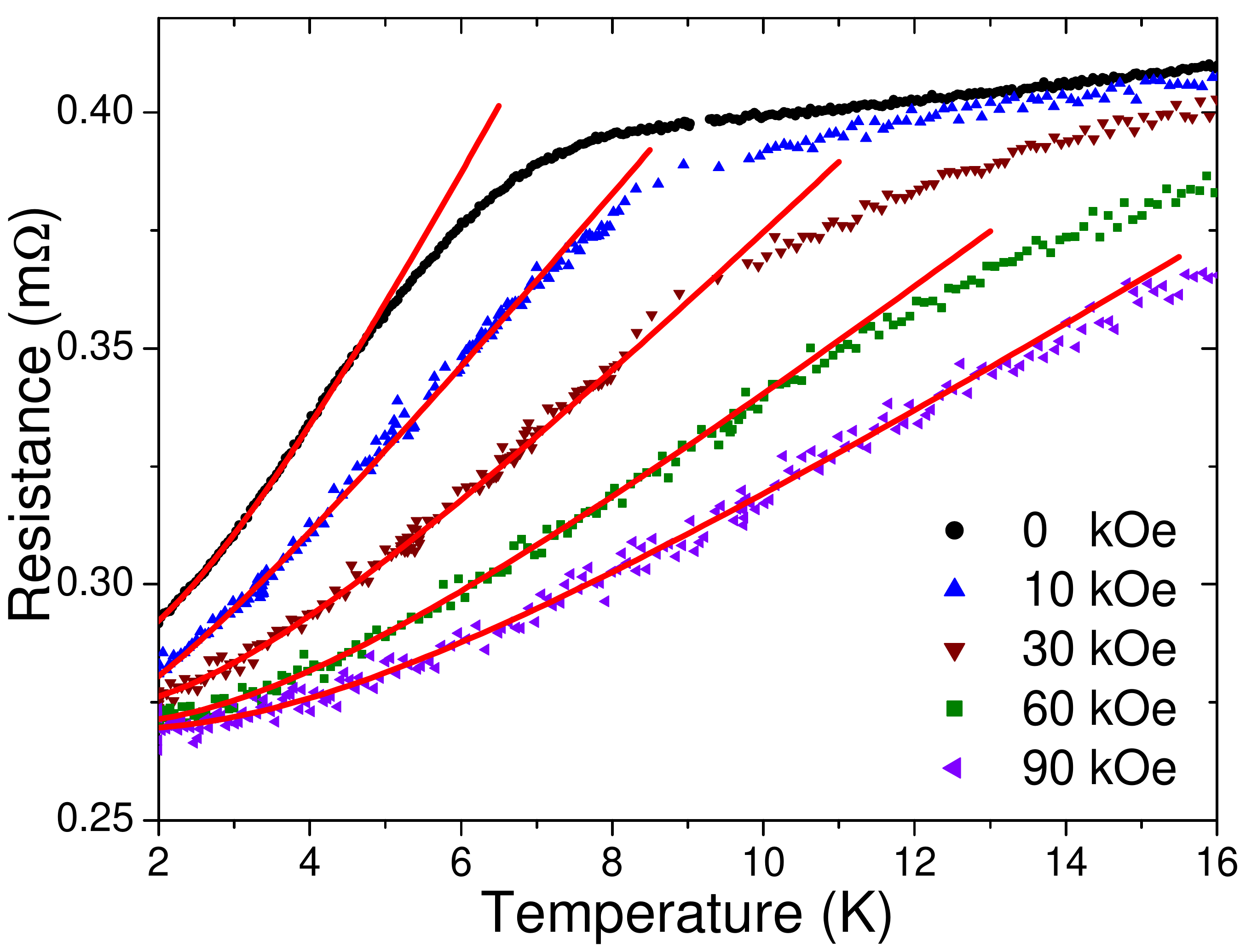}
\caption{\label{RTHf} Electrical resistance of \ce{YbNi_2} as a function of temperature at different magnetic fields indicated in the plot. The continuous lines are data fit using Eq. (2). }
\end{center}
\end{figure}

To analyze the electrical resistance of \ce{YbNi_2} below the
ferromagnetic transition temperature we assume that the main
scattering process is electron-magnon. With this assumption, we use
the Andersen and Smith model \cite{Andersen}. According to this
model, the magnetic resistance ($R_{mag}$) can be describe by:
\begin{equation}
 R_{mag}=AT\Delta e^{(-\Delta/T)} \left[1+2\frac{T}{\Delta}+\frac{1}{2}e^{(-\Delta/T)}+\cdots\right],
 \label{magnon1}
\end{equation}
where $\Delta$ is an energy gap in the magnonic density of states
and $A$ is a constant dependent on the material \cite{Andersen}. In
the case of $\Delta=0$ we deal with an isotropic ferromagnet and
then $R_{mag} \propto T^2$. However, fitting the experimental data
to a $T^2$ dependence the results were not acceptable. Taking in
account Eq. \ref{magnon1}, just with the first term, and considering
the residual resistance the data fit is good. Then the electrical
resistance is given by:
\begin{equation}
 R(T)=R_0+AT\Delta e^{(-\Delta/T)},
 \label{magnon}
\end{equation}
where $R_0$ is the residual resistance. We fit the $R(T,H)$ curves
at temperatures between 2 K and approximately 7.5 K. The phonon
contribution is considered negligible because the Curie temperature
$T_C=9$ K is small compared to the Debye temperature reported,
$\theta_D=272$ K \cite{Rojas}. Since the $R(T,H)$ curves below $T_C$
can be fitted with Eq. \ref{magnon}, we conclude that the same
scattering mechanism persists for the applied fields, i.e., the
electrical resistance is produced by the electron-magnon scattering.
Figure \ref{RTHf} shows the temperature dependence of the electrical
resistance in \ce{YbNi_2} at different magnetic fields in a
temperature range between 2 K and 16 K. In this figure the
continuous lines are the best fit obtained with Eq. \ref{magnon}.
Table \ref{magnonfit} presents the parameters obtained from the
fitting.

\begin{table}[ht]
 \caption{Parameters obtained from fits using Eq. 2. $R_0$ is the residual resistance, A is a constant related to material and $\Delta$ is the magnonic energy gap.}
 \begin{ruledtabular}
  \begin{tabular}{llll}
   H(kOe)&$R_0$(m$\Omega$)&A(m$\Omega$/K meV)&$\Delta$ (meV)\\
  \noalign{\smallskip}\hline\noalign{\smallskip}
   0  & 0.2817(5) &0.0089(1) & 0.31(1)\\
   10 & 0.2690(2) &0.0080(7) & 0.21(2)\\
   30 & 0.2730(6) &0.0035(1) & 0.39(3)\\
   60 & 0.2701(6) &0.0021(1) & 0.51(4)\\
   90 & 0.2691(6) &0.0014(1) & 0.62(8)\\
  \end{tabular}
  \label{magnonfit}
\end{ruledtabular}
\end{table}

\begin{figure}[ht]
\begin{center}
\includegraphics[width=0.48\textwidth]{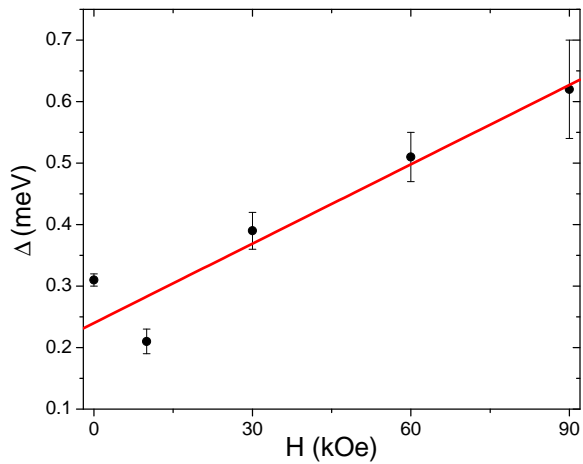}
\caption{\label{DvsB}Magnonic energy gap $\Delta$ as a function of the applied magnetic field. $\Delta$ was obtained from the fits of $R(T,H)$ curves of \ce{YbNi_2}. The continuous line is a linear fit of data.}
\end{center}
\end{figure}

It is well known that the magnonic energy gap is modified in the
presence of an external magnetic field as a consequence of the
Zeeman effect. The magnonic energy gap as a function of a magnetic
field is given by $\Delta=\Delta_{0}+g_J\mu_B \mu_0H$ \cite{Fontes},
where $\Delta_{0}$ is the magnonic energy gap at zero field,
$g_J=1.14$ is the Land\'e factor of $\rm{Yb^{3+}}$ ion and
$g_J\mu_B=0.065$ meVT$^{-1}$. The $g_J\mu_B$ value is the factor
that determines the change of $\Delta$ as a function of the magnetic
field. Figure \ref{DvsB} shows the magnonic energy gap as a function
of the applied magnetic field, as expected, $\Delta$ increases as
the magnetic field is increased. The best linear fit of $\Delta(H)$
gives that $\Delta(H)=0.24(4)+0.0043(6)H$, the slope must be
equivalent to $g_J\mu_B$, but its value is 0.043 meVT$^{-1}$ lower
than the calculated for the \ce{Yb^{3+}} ion. In addition, it was noted that the parameter $A$ depends on magnetic field.

\begin{figure}[ht]
\begin{center}
\includegraphics[width=0.48\textwidth]{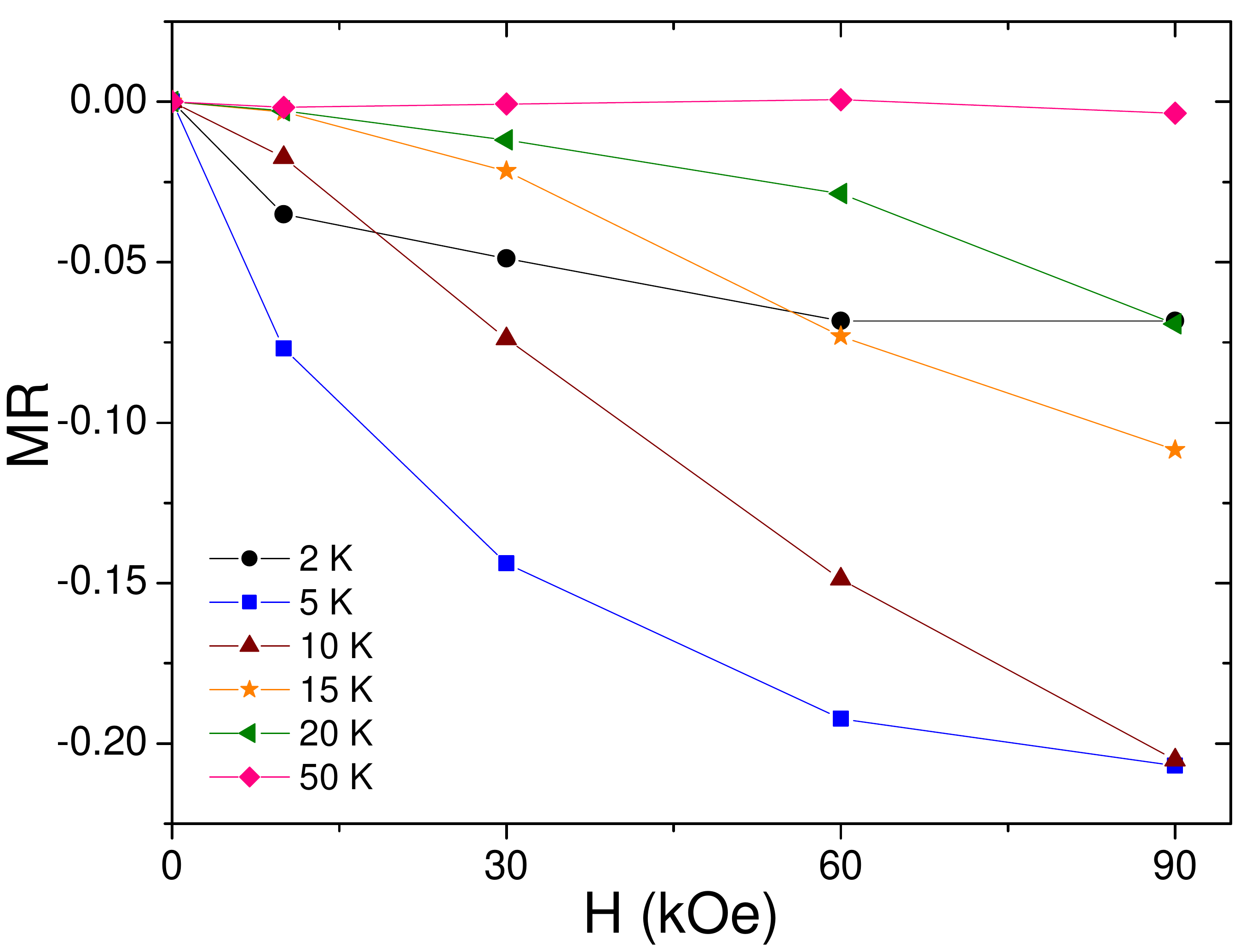}
\caption{\label{MR} Magnetoresistance $MR$ of \ce{YbNi_2} as a function of applied magnetic field at various temperatures. The data were extracted from curves of Fig. 3. The lines are a guide for the eye.}
\end{center}
\end{figure}

Finally, the magnetoresistance as a function of applied magnetic
field at several temperatures was extracted from the $R(T,H)$
curves (Fig. \ref{RTB}). The magnetoresistance was defined as
$MR=[R(H)-R(0)]/R(0)$, where $R(0)$ is the electrical resistance
measured without applied magnetic field and $R(H)$ is the electrical
resistance measured with an applied magnetic field $H$. Figure
\ref{MR} shows the plot of $MR(H)$ at 2, 5, 10, 15, 20 and 50 K. The
maximum $MR$ observed in \ce{YbNi_2} is about -20\% at 5 K and 10 K
(near $T_C$) under 90 kOe. At temperatures below 50 K, the $MR$ is
negative above and below $T_C$ for all applied magnetic fields. At
fixed magnetic field the $MR$ magnitude increases with temperature
until about $T_C$, for temperatures above $T_C$ it diminishes. This
behavior has been observed in other ferromagnetic Yb-based HF such
as \ce{YbPtGe} \cite{YbPtGe} and \ce{YbPdSi} \cite{YbPdSi}.  A
negative MR for $T<T_C$ can be associated with a reduction of the
spin-disorder suppressed by the magnetic field. For $T>T_C$, a
negative MR could be associated with the existence of ferromagnetic
short-range correlations or spin fluctuations that preceded the
onset of long-range magnetic order. Further experiments on
\ce{YbNi_2}, under magnetic field and/or pressure, must be performed
to get more insight in their physical properties.

\section{Conclusions}

We have synthesized polycrystalline samples of \ce{YbNi_2}, the
structure type and cell parameter were determined using X-ray
diffraction analysis. Magnetic measurement as a function of
temperature shows a ferromagnetic transition at 9 K. The effective
magnetic moment determined from the magnetic measurements suggest a
localized magnetism associated to \ce{Yb^{3+}} ions. Electrical
resistance as a function of temperature shows metallic-like behavior
at high temperatures. From this measurement, we infer that the RKKY
is the dominant interaction regard to Kondo effect. Below the
ferromagnetic transition temperature, the electrical resistance
results from the electron-magnon scattering, even in the presence of
an external magnetic field. In addition, the electrical resistance
analysis indicates the presence of an energy gap in the magnonic
spectrum indicating that the ferromagnetic state is anisotropic. The
increment of the energy gap can be associated to the Zeeman effect.

\begin{acknowledgments}
F.M. acknowledges the financial support from DGAPA-UNAM project
IN105917 and to R. Escudero by facilities to perform the magnetic
measurements. O.O. tanks the CONACyT scholarship.
\end{acknowledgments}

\bibliography{referenciasYbNi2}

\begin{thebibliography}{28}%
\makeatletter
\providecommand \@ifxundefined [1]{%
 \@ifx{#1\undefined}
}%
\providecommand \@ifnum [1]{%
 \ifnum #1\expandafter \@firstoftwo
 \else \expandafter \@secondoftwo
 \fi
}%
\providecommand \@ifx [1]{%
 \ifx #1\expandafter \@firstoftwo
 \else \expandafter \@secondoftwo
 \fi
}%
\providecommand \natexlab [1]{#1}%
\providecommand \enquote  [1]{``#1''}%
\providecommand \bibnamefont  [1]{#1}%
\providecommand \bibfnamefont [1]{#1}%
\providecommand \citenamefont [1]{#1}%
\providecommand \href@noop [0]{\@secondoftwo}%
\providecommand \href [0]{\begingroup \@sanitize@url \@href}%
\providecommand \@href[1]{\@@startlink{#1}\@@href}%
\providecommand \@@href[1]{\endgroup#1\@@endlink}%
\providecommand \@sanitize@url [0]{\catcode `\\12\catcode `\$12\catcode
  `\&12\catcode `\#12\catcode `\^12\catcode `\_12\catcode `\%12\relax}%
\providecommand \@@startlink[1]{}%
\providecommand \@@endlink[0]{}%
\providecommand \url  [0]{\begingroup\@sanitize@url \@url }%
\providecommand \@url [1]{\endgroup\@href {#1}{\urlprefix }}%
\providecommand \urlprefix  [0]{URL }%
\providecommand \Eprint [0]{\href }%
\providecommand \doibase [0]{http://dx.doi.org/}%
\providecommand \selectlanguage [0]{\@gobble}%
\providecommand \bibinfo  [0]{\@secondoftwo}%
\providecommand \bibfield  [0]{\@secondoftwo}%
\providecommand \translation [1]{[#1]}%
\providecommand \BibitemOpen [0]{}%
\providecommand \bibitemStop [0]{}%
\providecommand \bibitemNoStop [0]{.\EOS\space}%
\providecommand \EOS [0]{\spacefactor3000\relax}%
\providecommand \BibitemShut  [1]{\csname bibitem#1\endcsname}%
\let\auto@bib@innerbib\@empty
%</preamble>
\bibitem [{\citenamefont {Stewart}(2001)}]{Stewartd-f}%
  \BibitemOpen
  \bibfield  {author} {\bibinfo {author} {\bibfnamefont {G.~R.}\ \bibnamefont
  {Stewart}},\ }\href@noop {} {\bibfield  {journal} {\bibinfo  {journal} {Rev.
  Mod. Phys.}\ }\textbf {\bibinfo {volume} {73}},\ \bibinfo {pages} {797}
  (\bibinfo {year} {2001})}\BibitemShut {NoStop}%
\bibitem [{\citenamefont {Wirth}\ and\ \citenamefont {Steglich}(2016)}]{HF}%
  \BibitemOpen
  \bibfield  {author} {\bibinfo {author} {\bibfnamefont {S.}~\bibnamefont
  {Wirth}}\ and\ \bibinfo {author} {\bibfnamefont {F.}~\bibnamefont
  {Steglich}},\ }\href@noop {} {\bibfield  {journal} {\bibinfo  {journal}
  {Nature Reviews Materials}\ }\textbf {\bibinfo {volume} {1}},\ \bibinfo
  {pages} {16051} (\bibinfo {year} {2016})}\BibitemShut {NoStop}%
\bibitem [{\citenamefont {Stewart}(1984)}]{StewartHF}%
  \BibitemOpen
  \bibfield  {author} {\bibinfo {author} {\bibfnamefont {G.~R.}\ \bibnamefont
  {Stewart}},\ }\href@noop {} {\bibfield  {journal} {\bibinfo  {journal} {Rev.
  Mod. Phys.}\ }\textbf {\bibinfo {volume} {56}},\ \bibinfo {pages} {755}
  (\bibinfo {year} {1984})}\BibitemShut {NoStop}%
\bibitem [{\citenamefont {Bauer}(1991)}]{BauerCeYb}%
  \BibitemOpen
  \bibfield  {author} {\bibinfo {author} {\bibfnamefont {E.}~\bibnamefont
  {Bauer}},\ }\href@noop {} {\bibfield  {journal} {\bibinfo  {journal} {{Adv.
  Phys.}}\ }\textbf {\bibinfo {volume} {40}},\ \bibinfo {pages} {417} (\bibinfo
  {year} {1991})}\BibitemShut {NoStop}%
\bibitem [{\citenamefont {Varma}(1976)}]{MValence}%
  \BibitemOpen
  \bibfield  {author} {\bibinfo {author} {\bibfnamefont {C.~M.}\ \bibnamefont
  {Varma}},\ }\href@noop {} {\bibfield  {journal} {\bibinfo  {journal} {Rev.
  Mod. Phys.}\ }\textbf {\bibinfo {volume} {48}},\ \bibinfo {pages} {219}
  (\bibinfo {year} {1976})}\BibitemShut {NoStop}%
\bibitem [{\citenamefont {Ruderman}\ and\ \citenamefont
  {Kittel}(1954)}]{RKKY1}%
  \BibitemOpen
  \bibfield  {author} {\bibinfo {author} {\bibfnamefont {M.~A.}\ \bibnamefont
  {Ruderman}}\ and\ \bibinfo {author} {\bibfnamefont {C.}~\bibnamefont
  {Kittel}},\ }\href@noop {} {\bibfield  {journal} {\bibinfo  {journal} {Phys.
  Rev.}\ }\textbf {\bibinfo {volume} {96}},\ \bibinfo {pages} {99} (\bibinfo
  {year} {1954})}\BibitemShut {NoStop}%
\bibitem [{\citenamefont {Yosida}(1957)}]{RKKY2}%
  \BibitemOpen
  \bibfield  {author} {\bibinfo {author} {\bibfnamefont {Y.}~\bibnamefont
  {Yosida}},\ }\href@noop {} {\bibfield  {journal} {\bibinfo  {journal} {Phys.
  Rev.}\ }\textbf {\bibinfo {volume} {106}},\ \bibinfo {pages} {893} (\bibinfo
  {year} {1957})}\BibitemShut {NoStop}%
\bibitem [{\citenamefont {Gegenwart}\ \emph {et~al.}(2008)\citenamefont
  {Gegenwart}, \citenamefont {Si},\ and\ \citenamefont {Steglich}}]{QFT1}%
  \BibitemOpen
  \bibfield  {author} {\bibinfo {author} {\bibfnamefont {P.}~\bibnamefont
  {Gegenwart}}, \bibinfo {author} {\bibfnamefont {Q.}~\bibnamefont {Si}}, \
  and\ \bibinfo {author} {\bibfnamefont {F.}~\bibnamefont {Steglich}},\
  }\href@noop {} {\bibfield  {journal} {\bibinfo  {journal} {Nature Phys.}\
  }\textbf {\bibinfo {volume} {4}},\ \bibinfo {pages} {186} (\bibinfo {year}
  {2008})}\BibitemShut {NoStop}%
\bibitem [{\citenamefont {Si}\ and\ \citenamefont {Steglich}(2010)}]{QFT2}%
  \BibitemOpen
  \bibfield  {author} {\bibinfo {author} {\bibfnamefont {Q.}~\bibnamefont
  {Si}}\ and\ \bibinfo {author} {\bibfnamefont {F.}~\bibnamefont {Steglich}},\
  }\href@noop {} {\bibfield  {journal} {\bibinfo  {journal} {Science}\ }\textbf
  {\bibinfo {volume} {329}},\ \bibinfo {pages} {1161} (\bibinfo {year}
  {2010})}\BibitemShut {NoStop}%
\bibitem [{\citenamefont {Skrabek}\ and\ \citenamefont
  {Wallace}(1963)}]{skrabek}%
  \BibitemOpen
  \bibfield  {author} {\bibinfo {author} {\bibfnamefont {E.~A.}\ \bibnamefont
  {Skrabek}}\ and\ \bibinfo {author} {\bibfnamefont {W.~E.}\ \bibnamefont
  {Wallace}},\ }\href@noop {} {\bibfield  {journal} {\bibinfo  {journal} {J.
  Appl. Phys.}\ }\textbf {\bibinfo {volume} {34}},\ \bibinfo {pages} {1356}
  (\bibinfo {year} {1963})}\BibitemShut {NoStop}%
\bibitem [{\citenamefont {Wallace}\ \emph {et~al.}(1977)\citenamefont
  {Wallace}, \citenamefont {Sankar},\ and\ \citenamefont {Rao}}]{RNi2}%
  \BibitemOpen
  \bibfield  {author} {\bibinfo {author} {\bibfnamefont {W.~E.}\ \bibnamefont
  {Wallace}}, \bibinfo {author} {\bibfnamefont {S.~G.}\ \bibnamefont {Sankar}},
  \ and\ \bibinfo {author} {\bibfnamefont {V.~U.~S.}\ \bibnamefont {Rao}},\
  }in\ \href@noop {} {\emph {\bibinfo {booktitle} {New Concepts}}}\ (\bibinfo
  {publisher} {Springer Berlin Heidelberg},\ \bibinfo {address} {Berlin,
  Heidelberg},\ \bibinfo {year} {1977})\ pp.\ \bibinfo {pages}
  {1--55}\BibitemShut {NoStop}%
\bibitem [{\citenamefont {Rojas}\ \emph {et~al.}(2012)\citenamefont {Rojas},
  \citenamefont {{Fern{\'a}ndez Barqu{\'i}n}}, \citenamefont
  {Echevarria-Bonet},\ and\ \citenamefont {{Rodr{\'i}guez
  Fern{\'a}ndez}}}]{Rojas}%
  \BibitemOpen
  \bibfield  {author} {\bibinfo {author} {\bibfnamefont {D.~P.}\ \bibnamefont
  {Rojas}}, \bibinfo {author} {\bibfnamefont {L.}~\bibnamefont {{Fern{\'a}ndez
  Barqu{\'i}n}}}, \bibinfo {author} {\bibfnamefont {C.}~\bibnamefont
  {Echevarria-Bonet}}, \ and\ \bibinfo {author} {\bibfnamefont
  {J.}~\bibnamefont {{Rodr{\'i}guez Fern{\'a}ndez}}},\ }\href@noop {}
  {\bibfield  {journal} {\bibinfo  {journal} {Solid State Commun.}\ }\textbf
  {\bibinfo {volume} {152}},\ \bibinfo {pages} {1834} (\bibinfo {year}
  {2012})}\BibitemShut {NoStop}%
\bibitem [{\citenamefont {Yan}\ and\ \citenamefont {Wu}(2014)}]{yan14}%
  \BibitemOpen
  \bibfield  {author} {\bibinfo {author} {\bibfnamefont {E.}~\bibnamefont
  {Yan}}\ and\ \bibinfo {author} {\bibfnamefont {B.-N.}\ \bibnamefont {Wu}},\
  }\href@noop {} {\bibfield  {journal} {\bibinfo  {journal} {J. Supercond. Nov.
  Magn.}\ }\textbf {\bibinfo {volume} {27}},\ \bibinfo {pages} {735} (\bibinfo
  {year} {2014})}\BibitemShut {NoStop}%
\bibitem [{\citenamefont {Lutterotti}\ and\ \citenamefont
  {Gialanella}(1997)}]{maud}%
  \BibitemOpen
  \bibfield  {author} {\bibinfo {author} {\bibfnamefont {L.}~\bibnamefont
  {Lutterotti}}\ and\ \bibinfo {author} {\bibfnamefont {S.}~\bibnamefont
  {Gialanella}},\ }\href@noop {} {\bibfield  {journal} {\bibinfo  {journal}
  {Acta Mater.}\ }\textbf {\bibinfo {volume} {46}},\ \bibinfo {pages} {101}
  (\bibinfo {year} {1997})}\BibitemShut {NoStop}%
\bibitem [{\citenamefont {Buschow}(1972)}]{buschow72}%
  \BibitemOpen
  \bibfield  {author} {\bibinfo {author} {\bibfnamefont {K.~H.~J.}\
  \bibnamefont {Buschow}},\ }\href@noop {} {\bibfield  {journal} {\bibinfo
  {journal} {J. Less Commun Metals}\ }\textbf {\bibinfo {volume} {26}},\
  \bibinfo {pages} {329} (\bibinfo {year} {1972})}\BibitemShut {NoStop}%
\bibitem [{\citenamefont {Palenzona}\ and\ \citenamefont
  {Cirafici}(1973)}]{Palenzona}%
  \BibitemOpen
  \bibfield  {author} {\bibinfo {author} {\bibfnamefont {A.}~\bibnamefont
  {Palenzona}}\ and\ \bibinfo {author} {\bibfnamefont {S.}~\bibnamefont
  {Cirafici}},\ }\href@noop {} {\bibfield  {journal} {\bibinfo  {journal} {J.
  Less-Common Metals}\ }\textbf {\bibinfo {volume} {33}},\ \bibinfo {pages}
  {361} (\bibinfo {year} {1973})}\BibitemShut {NoStop}%
\bibitem [{\citenamefont {Ivanshin}\ \emph {et~al.}(2017)\citenamefont
  {Ivanshin}, \citenamefont {Gataullin}, \citenamefont {Sukhanov},
  \citenamefont {Ivanshin}, \citenamefont {Rojas},\ and\ \citenamefont
  {{Fern\'andez Barqu\'in}}}]{ivanshin14}%
  \BibitemOpen
  \bibfield  {author} {\bibinfo {author} {\bibfnamefont {V.~A.}\ \bibnamefont
  {Ivanshin}}, \bibinfo {author} {\bibfnamefont {E.~M.}\ \bibnamefont
  {Gataullin}}, \bibinfo {author} {\bibfnamefont {A.~A.}\ \bibnamefont
  {Sukhanov}}, \bibinfo {author} {\bibfnamefont {N.~A.}\ \bibnamefont
  {Ivanshin}}, \bibinfo {author} {\bibfnamefont {D.~P.}\ \bibnamefont {Rojas}},
  \ and\ \bibinfo {author} {\bibfnamefont {L.}~\bibnamefont {{Fern\'andez
  Barqu\'in}}},\ }\href@noop {} {\bibfield  {journal} {\bibinfo  {journal}
  {Phys. Metals Metallogr.}\ }\textbf {\bibinfo {volume} {118}},\ \bibinfo
  {pages} {341} (\bibinfo {year} {2017})}\BibitemShut {NoStop}%
\bibitem [{\citenamefont {Adachi}\ \emph {et~al.}(2005)\citenamefont {Adachi},
  \citenamefont {Imanaka},\ and\ \citenamefont {Kang}}]{adachi}%
  \BibitemOpen
  \bibinfo {editor} {\bibfnamefont {G.}~\bibnamefont {Adachi}}, \bibinfo
  {editor} {\bibfnamefont {N.}~\bibnamefont {Imanaka}}, \ and\ \bibinfo
  {editor} {\bibfnamefont {Z.~C.}\ \bibnamefont {Kang}},\ eds.,\ \href@noop {}
  {\emph {\bibinfo {title} {Binary rare-earth oxides}}}\ (\bibinfo  {publisher}
  {Springer},\ \bibinfo {year} {2005})\BibitemShut {NoStop}%
\bibitem [{\citenamefont {Blundell}(2001)}]{Blundell}%
  \BibitemOpen
  \bibfield  {author} {\bibinfo {author} {\bibfnamefont {S.}~\bibnamefont
  {Blundell}},\ }\href@noop {} {\emph {\bibinfo {title} {{Magnetism in
  Condensed Matter}}}}\ (\bibinfo  {publisher} {Oxford University Press},\
  \bibinfo {year} {2001})\BibitemShut {NoStop}%
\bibitem [{\citenamefont {Blanco}\ \emph {et~al.}(1994)\citenamefont {Blanco},
  \citenamefont {de~Podesta}, \citenamefont {Espeso}, \citenamefont {G\'omez},
  \citenamefont {Lester}, \citenamefont {McEewn}, \citenamefont {Patrikios},\
  and\ \citenamefont {{Rodri\'iguez Fern\'andez}}}]{blanco94}%
  \BibitemOpen
  \bibfield  {author} {\bibinfo {author} {\bibfnamefont {J.~A.}\ \bibnamefont
  {Blanco}}, \bibinfo {author} {\bibfnamefont {M.}~\bibnamefont {de~Podesta}},
  \bibinfo {author} {\bibfnamefont {J.~I.}\ \bibnamefont {Espeso}}, \bibinfo
  {author} {\bibfnamefont {J.~C.}\ \bibnamefont {G\'omez}}, \bibinfo {author}
  {\bibfnamefont {C.}~\bibnamefont {Lester}}, \bibinfo {author} {\bibfnamefont
  {K.~A.}\ \bibnamefont {McEewn}}, \bibinfo {author} {\bibfnamefont
  {N.}~\bibnamefont {Patrikios}}, \ and\ \bibinfo {author} {\bibfnamefont
  {J.}~\bibnamefont {{Rodri\'iguez Fern\'andez}}},\ }\href@noop {} {\bibfield
  {journal} {\bibinfo  {journal} {Phys. Rev. B}\ }\textbf {\bibinfo {volume}
  {49}},\ \bibinfo {pages} {15126} (\bibinfo {year} {1994})}\BibitemShut
  {NoStop}%
\bibitem [{\citenamefont {Yamauchi}\ and\ \citenamefont
  {Fukamichi}(2000)}]{yamauchi00}%
  \BibitemOpen
  \bibfield  {author} {\bibinfo {author} {\bibfnamefont {R.}~\bibnamefont
  {Yamauchi}}\ and\ \bibinfo {author} {\bibfnamefont {K.}~\bibnamefont
  {Fukamichi}},\ }\href@noop {} {\bibfield  {journal} {\bibinfo  {journal} {J.
  Phys.: Condens. Matter}\ }\textbf {\bibinfo {volume} {12}},\ \bibinfo {pages}
  {2461} (\bibinfo {year} {2000})}\BibitemShut {NoStop}%
\bibitem [{\citenamefont {Kotur}\ \emph {et~al.}(2010)\citenamefont {Kotur},
  \citenamefont {Myakush}, \citenamefont {Michor},\ and\ \citenamefont
  {Bauer}}]{RRR-RNi2}%
  \BibitemOpen
  \bibfield  {author} {\bibinfo {author} {\bibfnamefont {R.}~\bibnamefont
  {Kotur}}, \bibinfo {author} {\bibfnamefont {O.}~\bibnamefont {Myakush}},
  \bibinfo {author} {\bibfnamefont {H.}~\bibnamefont {Michor}}, \ and\ \bibinfo
  {author} {\bibfnamefont {E.}~\bibnamefont {Bauer}},\ }\href@noop {}
  {\bibfield  {journal} {\bibinfo  {journal} {J. Alloys Compd.}\ }\textbf
  {\bibinfo {volume} {499}},\ \bibinfo {pages} {135} (\bibinfo {year}
  {2010})}\BibitemShut {NoStop}%
\bibitem [{\citenamefont {Gratz}\ \emph {et~al.}(1996)\citenamefont {Gratz},
  \citenamefont {Kottar}, \citenamefont {Lindbaum}, \citenamefont {Mantler},
  \citenamefont {Latroche}, \citenamefont {Paul-Boncour}, \citenamefont {Acet},
  \citenamefont {Barner}, \citenamefont {Holzapfel}, \citenamefont {Pacheco},\
  and\ \citenamefont {Yvon}}]{gratz96}%
  \BibitemOpen
  \bibfield  {author} {\bibinfo {author} {\bibfnamefont {E.}~\bibnamefont
  {Gratz}}, \bibinfo {author} {\bibfnamefont {A.}~\bibnamefont {Kottar}},
  \bibinfo {author} {\bibfnamefont {A.}~\bibnamefont {Lindbaum}}, \bibinfo
  {author} {\bibfnamefont {M.}~\bibnamefont {Mantler}}, \bibinfo {author}
  {\bibfnamefont {M.}~\bibnamefont {Latroche}}, \bibinfo {author}
  {\bibfnamefont {V.}~\bibnamefont {Paul-Boncour}}, \bibinfo {author}
  {\bibfnamefont {M.}~\bibnamefont {Acet}}, \bibinfo {author} {\bibfnamefont
  {C.~I.}\ \bibnamefont {Barner}}, \bibinfo {author} {\bibfnamefont {W.~B.}\
  \bibnamefont {Holzapfel}}, \bibinfo {author} {\bibfnamefont {V.}~\bibnamefont
  {Pacheco}}, \ and\ \bibinfo {author} {\bibfnamefont {K.}~\bibnamefont
  {Yvon}},\ }\href@noop {} {\bibfield  {journal} {\bibinfo  {journal} {J.
  Phys.: Condens. Matter}\ }\textbf {\bibinfo {volume} {8}},\ \bibinfo {pages}
  {8351} (\bibinfo {year} {1996})}\BibitemShut {NoStop}%
\bibitem [{\citenamefont {Moriya}(1985)}]{Moriya}%
  \BibitemOpen
  \bibfield  {author} {\bibinfo {author} {\bibfnamefont {T.}~\bibnamefont
  {Moriya}},\ }\href@noop {} {\emph {\bibinfo {title} {Spin fluctuations in
  itinerant electron magnetism}}},\ Springer series in solid-state sciences\
  (\bibinfo  {publisher} {Springer-Verlag},\ \bibinfo {year}
  {1985})\BibitemShut {NoStop}%
\bibitem [{\citenamefont {Andersen}\ and\ \citenamefont
  {Smith}(1979)}]{Andersen}%
  \BibitemOpen
  \bibfield  {author} {\bibinfo {author} {\bibfnamefont {N.~H.}\ \bibnamefont
  {Andersen}}\ and\ \bibinfo {author} {\bibfnamefont {H.}~\bibnamefont
  {Smith}},\ }\href@noop {} {\bibfield  {journal} {\bibinfo  {journal} {Phys.
  Rev. B}\ }\textbf {\bibinfo {volume} {19}},\ \bibinfo {pages} {384} (\bibinfo
  {year} {1979})}\BibitemShut {NoStop}%
\bibitem [{\citenamefont {Fontes}\ \emph {et~al.}(1999)\citenamefont {Fontes},
  \citenamefont {Trochez}, \citenamefont {Giordanengo}, \citenamefont {Bud'ko},
  \citenamefont {Sanchez}, \citenamefont {Baggio-Saitovitch},\ and\
  \citenamefont {Continentino}}]{Fontes}%
  \BibitemOpen
  \bibfield  {author} {\bibinfo {author} {\bibfnamefont {M.~B.}\ \bibnamefont
  {Fontes}}, \bibinfo {author} {\bibfnamefont {J.~C.}\ \bibnamefont {Trochez}},
  \bibinfo {author} {\bibfnamefont {B.}~\bibnamefont {Giordanengo}}, \bibinfo
  {author} {\bibfnamefont {S.~L.}\ \bibnamefont {Bud'ko}}, \bibinfo {author}
  {\bibfnamefont {D.~R.}\ \bibnamefont {Sanchez}}, \bibinfo {author}
  {\bibfnamefont {E.~M.}\ \bibnamefont {Baggio-Saitovitch}}, \ and\ \bibinfo
  {author} {\bibfnamefont {M.~A.}\ \bibnamefont {Continentino}},\ }\href@noop
  {} {\bibfield  {journal} {\bibinfo  {journal} {Phys. Rev. B}\ }\textbf
  {\bibinfo {volume} {60}},\ \bibinfo {pages} {6781} (\bibinfo {year}
  {1999})}\BibitemShut {NoStop}%
\bibitem [{\citenamefont {Katoh}\ \emph {et~al.}(2010)\citenamefont {Katoh},
  \citenamefont {Koga}, \citenamefont {Terui},\ and\ \citenamefont
  {Ochiai}}]{YbPtGe}%
  \BibitemOpen
  \bibfield  {author} {\bibinfo {author} {\bibfnamefont {K.}~\bibnamefont
  {Katoh}}, \bibinfo {author} {\bibfnamefont {T.}~\bibnamefont {Koga}},
  \bibinfo {author} {\bibfnamefont {G.}~\bibnamefont {Terui}}, \ and\ \bibinfo
  {author} {\bibfnamefont {A.}~\bibnamefont {Ochiai}},\ }\href@noop {}
  {\bibfield  {journal} {\bibinfo  {journal} {J. Phys. Soc. Jpn.}\ }\textbf
  {\bibinfo {volume} {79}},\ \bibinfo {pages} {084709} (\bibinfo {year}
  {2010})}\BibitemShut {NoStop}%
\bibitem [{\citenamefont {Tsujii}\ \emph {et~al.}(2016)\citenamefont {Tsujii},
  \citenamefont {Keller}, \citenamefont {D\"{o}nni},\ and\ \citenamefont
  {Kitazawa}}]{YbPdSi}%
  \BibitemOpen
  \bibfield  {author} {\bibinfo {author} {\bibfnamefont {N.}~\bibnamefont
  {Tsujii}}, \bibinfo {author} {\bibfnamefont {L.}~\bibnamefont {Keller}},
  \bibinfo {author} {\bibfnamefont {A.}~\bibnamefont {D\"{o}nni}}, \ and\
  \bibinfo {author} {\bibfnamefont {H.}~\bibnamefont {Kitazawa}},\ }\href@noop
  {} {\bibfield  {journal} {\bibinfo  {journal} {Journal of Physics: Condens.
  Matter}\ }\textbf {\bibinfo {volume} {28}},\ \bibinfo {pages} {336002}
  (\bibinfo {year} {2016})}\BibitemShut {NoStop}%
\end{thebibliography}%

\end{document}